\documentclass[prd,twocolumn,nofootinbib,preprintnumbers,showpacs]{revtex4-1}

\usepackage{amsfonts}
\usepackage{amsmath}
\usepackage{amssymb}
\usepackage{bm}
\usepackage{graphicx}
\usepackage{graphics}
\usepackage{latexsym}
\usepackage{url}
\usepackage{xspace}
\usepackage[usenames]{color}
\usepackage{mathrsfs}
\usepackage{hyperref}
\usepackage{epstopdf}
\usepackage{tensor}
\usepackage{mathtools}
\usepackage{siunitx}

\newcommand{\be}{\begin{equation}}
\newcommand{\ee}{\end{equation}}
\newcommand{\bal}{\begin{align}}

\newcommand{\nn}{\nonumber}

\newcommand{\pd}{\partial}

\newcommand{\tn}{\tensor}

\newcommand{\ssqth}{\sin^2{\theta}}

\newcommand{\inter}{\mathrm{int}}
\newcommand{\mcl}{\mathcal}

\newcommand{\gr}{{\mbox{\tiny GR}}}
\newcommand{\g}{{\mbox{\tiny (g)}}}
\newcommand{\f}{{\mbox{\tiny (f)}}}
\newcommand{\mrm}{\mathrm}
\newcommand{\rarr}{\rightarrow}

\newcommand{\lb}{\left(}
\newcommand{\rb}{\right)}
\newcommand{\lcb}{\left\{}
\newcommand{\rcb}{\right\}}
\newcommand{\lsb}{\left[}
\newcommand{\rsb}{\right]}

\begin{document}
\title{Slowly-Rotating Neutron Stars in Massive Bigravity}

\author{Andrew Sullivan}
\author{Nicol\'as Yunes}
\affiliation{eXtreme Gravity Institute, Department of Physics, Montana State University, Bozeman, MT 59717, USA.}

\date{\today}

\begin{abstract} 

We study slowly-rotating neutron stars in ghost-free massive bigravity. 
This theory modifies General Relativity by introducing a second, auxiliary but dynamical tensor field that couples to matter through the physical metric tensor through non-linear interactions. 
We expand the field equations to linear order in slow rotation and numerically construct solutions in the interior and exterior of the star with a set of realistic equations of state. 
We calculate the physical mass function with respect to observer radius and find that, unlike in General Relativity, this function does not remain constant outside the star; rather, it asymptotes to a constant a distance away from the surface, whose magnitude is controlled by the ratio of gravitational constants.
The Vainshtein-like radius at which the physical and auxiliary mass functions asymptote to a constant is controlled by the graviton mass scaling parameter, and outside this radius, bigravity modifications are suppressed. 
We also calculate the frame-dragging metric function and find that bigravity modifications are typically small in the entire range of coupling parameters explored. 
We finally calculate both the mass-radius and the moment of inertia-mass relations for a wide range of coupling parameters and find that both the graviton mass scaling parameter and the ratio of the gravitational constants introduce large modifications to both. 
These results could be used to place  future constraints on bigravity with electromagnetic and gravitational-wave observations of isolated and binary neutron stars.
\end{abstract}

\maketitle

\section{Introduction}

In recent years, the exploration of the accelerated expansion of the universe has led to a renewed interest in modified gravity theories with a massive graviton. It was previously thought that any theory with a massive graviton would introduce an unphysical ghost-like scalar propagating mode, until it was shown in 2009~\cite{deRham:2010kj,deRham:2014zqa} that the ghost mode could be removed with a properly generalized action. In the resulting theory (dRGT or massive gravity), a second metric, taken to be flat and non-dynamical, was introduced in order to generate the non-trivial higher order terms in the action that were required to remove the ghost mode. In 2011, it was shown~\cite{Hassan:2011vm} that the dRGT theory could be generalized by removing the requirement that the auxiliary metric be non-dynamical. The resulting theory, aptly named massive bigravity or bimetric gravity~\cite{1751-8121-49-18-183001}, is one in which there are two metrics, a physical one that couples to matter and an auxiliary but dynamical one that couples directly to the metric.  

Most of the work thus far has been focused on cosmology in massive gravity and massive bigravity. Shortly after this renewed interest it was found \cite{DAmico:2011eto} that there are no spatially flat or closed Friedman-Lema\^itre-Robertson-Walker (FLRW) solutions in massive gravity due to a modified Hamiltonian constraint. Although there are cosmological solutions that can explain the present and past observations of the evolution of the universe, many of these solutions suffer from a fine-tuning problem, and many of these exact solutions have been found to be unstable. There have been multiple avenues to address these issues, one of which was to add additional degrees of freedom to the theory by allowing the auxiliary metric in massive gravity to become dynamical, which promoted the initial interest in bimetric gravity. Adding dynamics to the auxiliary metric places it on equal footing with the primary metric and the gravitational force in each metric is mediated by its own graviton, one being massless and one being massive. It was subsequently shown that massive bigravity does admit spatially open, closed, and flat cosmological solutions~\cite{Volkov:2011an}, although cosmology in this theory is still very much an ongoing area of research.

\begin{figure*}[htbp]
\begin{center}
\resizebox{8.75cm}{!}{\include{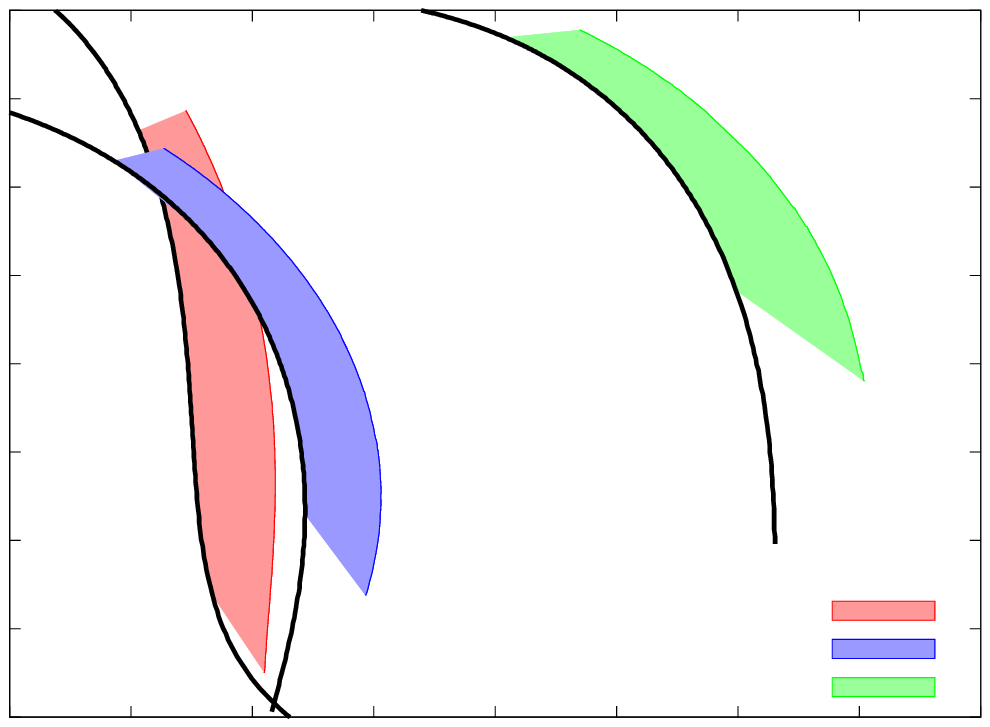}} \quad
\resizebox{8.75cm}{!}{\include{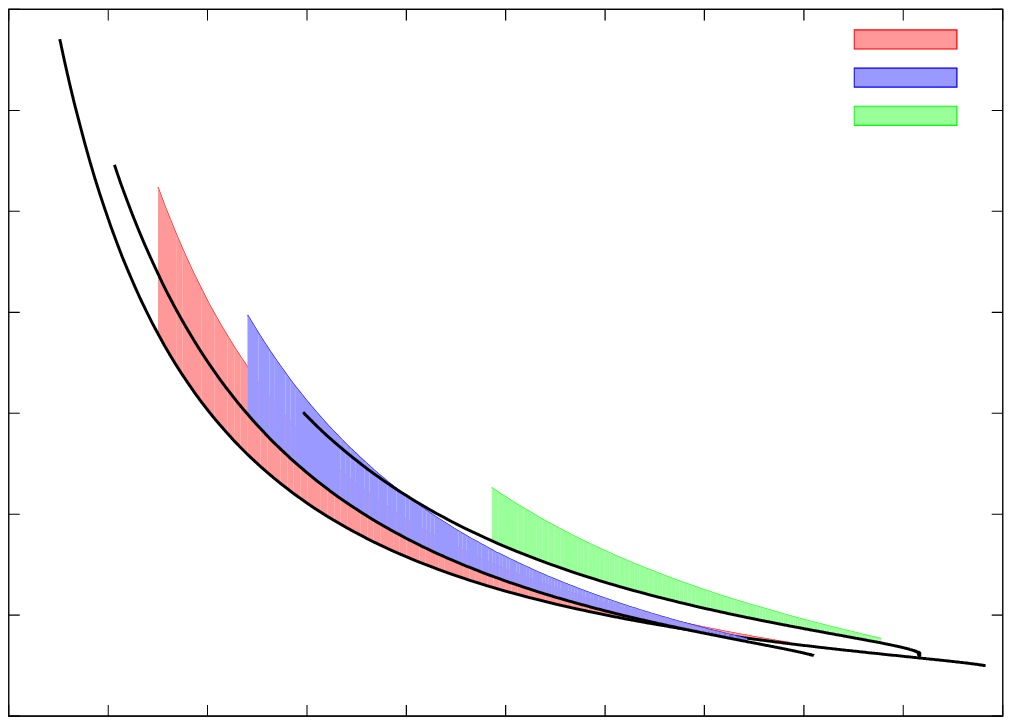}}
\caption{\label{fig:MR}~[Color Online]. The mass-radius relation (left) and the moment of inertia-mass relation (right) using three realistic equations of state with varying $m$ and all other coupling parameters fixed. The mass of the star $M$ is obtained from the mass function of the physical metric evaluated at the stellar surface defined by radius $R$. The dimensionless moment of inertia is defined via $\bar{I} = I/M^{3}$, where $I \equiv |\vec{J}|/\Omega$, with $\vec{J}$ the spin angular momentum and $\Omega$ the angular frequency of rotation. The shaded regions represent variability of the relations with the coupling parameter $m$ from $m \in [0, \num{3.0e-7} \mrm{cm}^{-1}]$, while keeping all others fixed to $c_3 = 0.48$, $c_4 = 1.71$, and $\eta = \pi/4$, with the solid black lines representing the GR limit, and the different colors representing different equations of state. Observe that the relations smoothly approach the GR relation in the $m \to 0$ limit. Observe also that as the mass of the star increases, the GR modification in the moment of inertia decreases, because the interior matter content of the star increasingly dominates the interaction with the auxiliary metric.}
\end{center}
\vspace{-0.5cm}
\end{figure*}

The existence and stability of black holes in massive bigravity is also an active area of research. The first spherically-symmetric, vacuum solutions in ghost-free massive bigravity emerged in 2011~\cite{Comelli:2011wq} and 2012~\cite{Volkov:2012wp}. These solutions were organized into two classes: bidiagonal (both metrics are diagonal) and non-bidiagonal (the physical metric is diagonal while the auxiliary one is not)~\cite{Comelli:2011wq}. For the bidiagonal class, perturbative solutions around a Minkowski background were found in~\cite{Volkov:2012wp}, as well as full non-linear solutions in~\cite{Brito:2013xaa} shortly after. The simplest case of bidiagonal vacuum solutions where both metrics are Schwarzschild-like is linearly unstable~\cite{Babichev:2013una}. For the non-bidiagonal class, solutions have been found analytically~\cite{Comelli:2011wq} and they have been shown to be stable~\cite{Babichev:2014oua}. The first study of axially-symmetric vacuum solutions showed the existence of rotating black holes~\cite{Babichev:2014tfa} and asymptotically de-Sitter rotating black holes~\cite{Ayon-Beato:2015qtt}. The simplest case of the former, where both metrics are Kerr-like, is linearly unstable~\cite{Brito:2013wya}.

The first treatment of non-rotating, spherically symmetric stars in massive bigravity was completed assuming a uniform density distribution in 2012~\cite{Volkov:2012wp} and 2015~\cite{1475-7516-2015-11-023}. These solutions successfully demonstrated the Vainshtein mechanism to recover GR~\cite{Volkov:2012wp}, which was then used to place constraints on the ratio of gravitational constants and the Compton wavelength of the graviton~\cite{1475-7516-2015-11-023}. Shortly after, non-rotating, spherically symmetric stars with matter satisfying a polytropic equation of state were found~\cite{Aoki:2016eov}, and were organized into two main classes determined by a critical compactness. Above this critical compactness no regular solutions were found, which leads to a maximum neutron star mass that is much smaller than that in GR. Given that more massive neutron stars have already been observed~\cite{Lorimer2008,Demorest:2010bx}, this class of solutions is observationally ruled out. Below this critical compactness, the Vainshtein mechanism is able to successfully recover GR, avoiding this strict mass limit. In this paper, we extend these analyses in two main ways: (i) we consider ``realistic'' nuclear equations of state, and (ii) we extend the analysis to slowly-rotating neutron stars.

We begin with the Hassan-Rosen action in massive bigravity~\cite{Hassan:2011vm}, which depends on two metrics: a ``physical'' g-metric that couples directly to the matter sector and an auxiliary, dynamical f-metric that couples directly only to the g-metric. The gravitational action is then simply two copies of the Einstein-Hilbert action (one for each metric), plus a set of metric interaction terms that couple the two metrics together. The couplings between metrics and between the g-metric and matter is controlled by various coupling parameters, of which only $m$, $c_3$, $c_4$, and $\eta$ will be relevant here. The parameter $\eta$ characterizes the relative strength of the gravitational constants ($G$ and $\mcl{G}$), which multiply the Einstein-Hilbert terms, and it is bounded by $\eta \in [0,\frac{\pi}{2}]$. The  parameter $m$ quantifies the overall strength of all metric interaction terms in the action, and it is closely related to the mass of the graviton in the theory. The parameters $c_3$ and $c_4$ quantify the relative strength of the high-order interaction terms in the action, and are scaled to be of order unity. Section~\ref{sec:MBG} provides a more detailed summary of the theory. 

We here study slowly-rotating neutron stars in isolation, and thus, we can simplify the field equations significantly. First, we require the physical and auxiliary metrics to be both stationary and axisymmetric with asymptotically flat boundary conditions at spatial infinity.  Second, we model the matter stress-energy tensor as a perfect fluid with a barotropic equation of state, obtained numerically from nuclear physics calculations; in particular, we consider the Akmal-Pandharipande-Ravenhall (APR)~\cite{Akmal:1998cf}, Lattimer-Swesty (LS220)~\cite{LATTIMER1991331}, and Shen et al.~\cite{Shen:1998gq} equations of state. Third, we solve the field equations in an expansion about small rotation, assuming that the product of the angular velocity and the radius of the star is much smaller than unity.  Finally, we simplify the field equations using a tetrad basis decomposition~\cite{Volkov:2012wp} that allows us to obtain modified Tolman-Oppenheimer-Volkoff (TOV) equations at zeroth-order in rotation, and a set of field equations at first-order in rotation.

The numerical evolution of these equations requires that we choose boundary conditions carefully. We carry out a local analysis of the field equations at the stellar core and at spatial infinity, which allows us to find power-series solutions in these two regimes. We then use these power-series expressions as boundary conditions for the numerical evolution of the field equations order by order in a slow-rotation expansion. We use a shooting method to match all metric functions and their first derivatives at the surface of the star, defined as the radius where the pressure vanishes, thus ensuring continuity and differentiability at the surface.

\begin{figure*}[htbp]
\begin{center}
\resizebox{8.75cm}{!}{\include{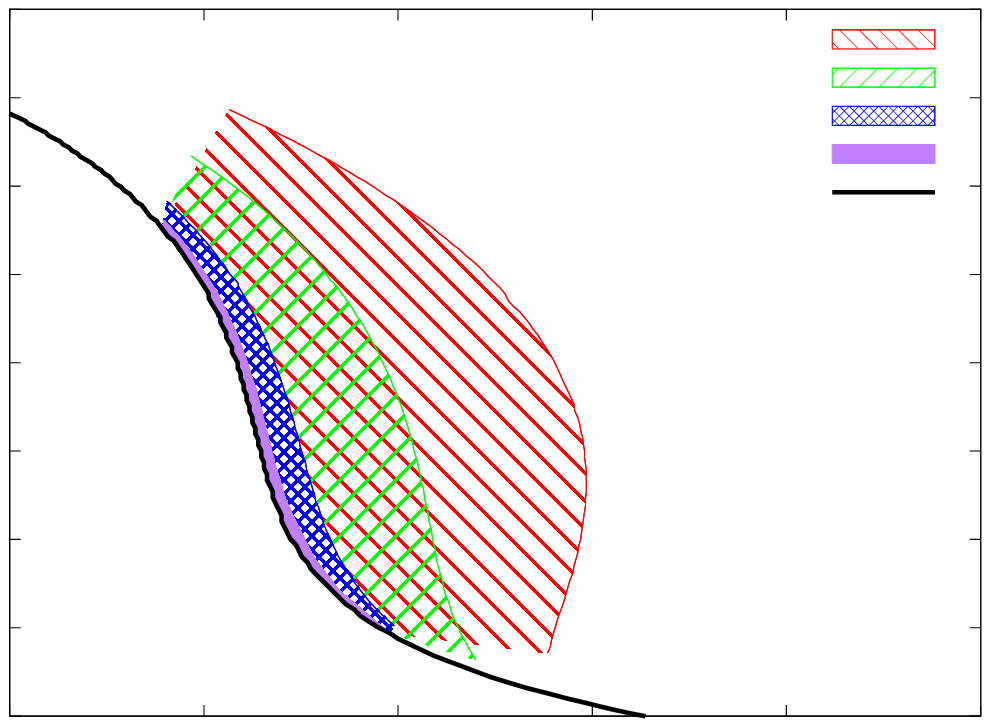}} \quad
\resizebox{8.75cm}{!}{\include{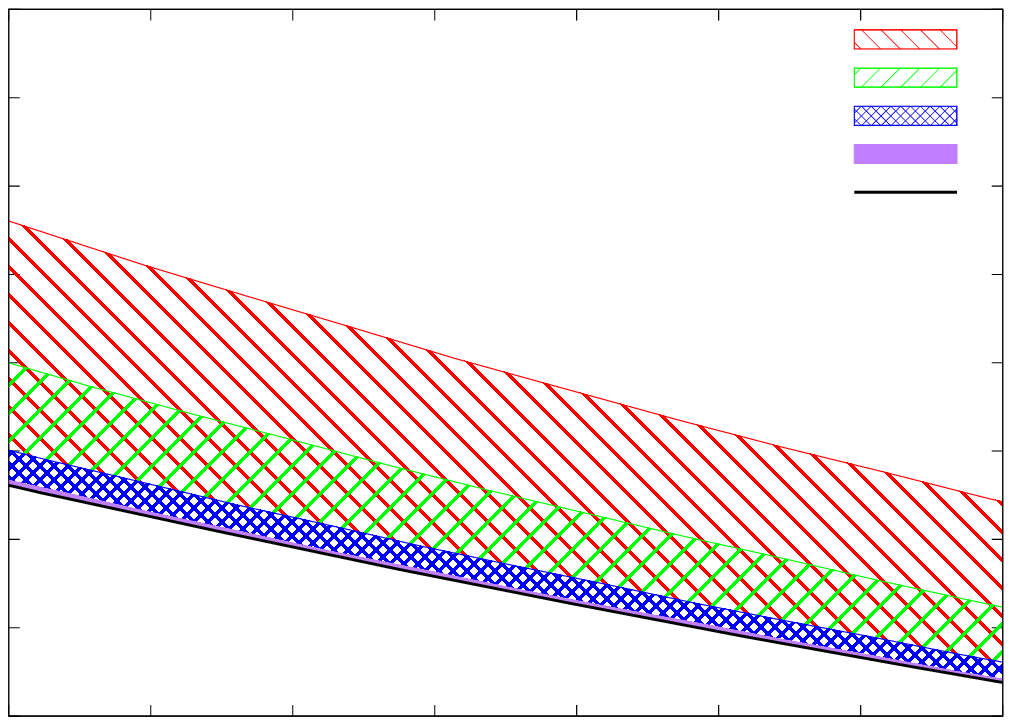}}
\caption{\label{fig:etaplot}~[Color Online]. Mass-radius (left) and moment of inertia-mass (right) relations for the APR equation of state for a set of fixed $m$ and varying $\eta \in (0, \pi/2)$. The lower bound on each region corresponds to the GR limit with $\eta = {\pi}/{2}$, while the upper bound corresponds to the $\eta = 0$ case and it is determined by the value of $m$. Observe that the fixed value of $m$ controls the maximum range of variability of these relations.}
\end{center}
\end{figure*}

From the zeroth-order modified TOV equations, we numerically generate spherically-symmetric stellar solutions with which we can calculate observables, such as the mass-radius relations shown in the left panel of Fig.~\ref{fig:MR}. As expected, the modified mass-radius relation in bigravity is highly degenerate with the equation of state, and it has a smooth transition to GR in both the $m \to 0$ and $\eta \rarr \pi/2$ limit. From the first-order field equations, we numerically generate the linear-in-spin corrections to the metric with which we can calculate other observables, such as the moment-of-inertia-mass curve shown in the right panel of Fig.~\ref{fig:MR}. As expected, the modified moment of inertia also has a smooth limit to GR, but in this case notice that as the moment of inertia increases, all curves tend to the GR limit. This is because as the central density increases, the mass increases and the matter interactions dominate over interactions between the auxiliary and physical metrics. 

The relations discussed above do not display a very interesting result in massive bigravity for neutron stars: the physical mass function outside the star does not remain constant as expected in GR, but rather it decreases due to interactions with the auxiliary metric (a type of Vainshtein screening). Inside the star, the physical mass $m_{g}$ increases much more rapidly than the auxiliary mass $m_f$ due to the explicit matter coupling to the physical metric. Outside the star, although the energy-density vanishes, the interaction between the physical and the auxiliary metrics causes the physical mass to decrease and asymptote to a constant that depends on the coupling parameter $\eta$. When $\eta = {\pi}/{4}$, both gravitational constants have equal strength and the asymptotic physical mass is exactly half of the Arnowit-Deser-Misner (ADM) mass in GR. As $\eta \to {\pi}/{2}$, the asymptotic physical mass becomes constant in the exterior spacetime, while as $\eta \to 0$, the asymptotic physical mass decreases until the neutron star mass is entirely screened by the auxiliary metric.

This behavior of the physical mass with $\eta$ means there are now two distinct limits to GR in this theory, when $m \to 0$ with $\eta$ fixed and when $\eta \to {\pi}/{2}$ with $m$ fixed, as shown in Fig.~\ref{fig:etaplot}. The left panel of this figure presents the mass-radius relation, while the right panel shows the moment of inertia-mass relation, for a single equation of state but varying both $m$ and $\eta$. In the $\eta \to 0$ case, the neutron star mass is entirely screened by the auxiliary metric, and as $m$ increases, the $\eta \to 0$ curve deviates more and more from GR. On the other hand, even for large values of $m$, when $\eta \sim {\pi}/{2}$, one still recovers GR.

We also explore the effects of the two remaining coupling parameters, $c_3$ and $c_4$ on the moment of inertia. We find that the moment of inertia is not substantially affected by these parameters within the range of values we considered, which are restricted by analytical considerations of scattering amplitudes in massive gravity~\cite{Cheung:2016yqr}. Unlike $m$ and $\eta$, the primary effect of both $c_3$ and $c_4$ appears to be to broaden the $I$-$M$ curve at each $m$ and $\eta$ value. As the mass scaling parameter $m$ increases, this broadening effect also increases. The weaker influence of $c_3$ and $c_4$ is due to the manner in which these parameters appear in the action. Since they couple to the third and fourth order terms of the action respectively, their effect is inherently going to be subdominant to the leading second order term scaled by $m$ and $\eta$, unless all interaction terms are non-linearly relevant. 

The remainder of this paper presents the details of the calculations and results presented above and it is organized as follows. In Sec.~\ref{sec:MBG}, we outline the basics of massive bigravity and introduce the notation we use throughout the paper. In Sec.~\ref{sec:nrBG}, we present the metric ansatz and the stress-energy tensor we employ, as well as the method to determine the interaction tensor in order to derive the modified TOV equations in massive bigravity. In Sec.~\ref{sec:rotBG}, we outline the asymptotic expansions used to obtain the boundary conditions for the numerical integrations, and describe the numerical algorithm we used to solve the modified field equations at both zeroth-order and first-order in spin. In Sec.~\ref{sec:conc}, we describe why the results obtained here are important and how one could use them to compare against observations of isolated and double binary pulsar systems. Throughout this paper, we will use the following conventions: Greek letters denote spacetime indices; capitalized Latin letters denote tetrad indices; both metrics have the spacetime signature $\lb -,+,+,+ \rb$; unless otherwise stated, geometric units where $G=c=1$.

\section{Massive Bigravity}
\label{sec:MBG}

The ghost-free action of massive bigravity given by Hassan and Rosen~\cite{Hassan:2011zd} in natural units ($c = \hbar$ = 1) is
\bal
\label{eq:action}
S &= M_g^2 \int R \sqrt{-g} \, d^4 x + M_f^2 \int \mcl{R} \sqrt{-f}\,d^4x \nn \\
& + 2 m^2 M_{\mathrm{eff}}^2 \int \mcl{L}_{\inter} \sqrt{-g} \, d^4 x + S_{\mrm{mat}} [\tn{g}{_\mu_\nu}],
\end{align}
where $R$ and $\mcl{R}$ are the Ricci scalars associated with the physical metric $g_{\mu \nu}$ and the auxiliary metric $f_{\mu \nu}$ respectively, with reduced Planck masses $M_g^2 = {\lb 8 \pi G \rb}^{-1}$ and $M_f^2 = {\lb 8 \pi \mcl{G} \rb}^{-1}$, and gravitational couplings $G$ and $\mcl{G}$. All matter degrees of freedom are encoded in $S_{\mrm{mat}}$, which couples directly only to the physical metric. 

The physical and auxiliary metrics communicate with each other through the interaction term $\mcl{L}_{\inter}$, which can be written as
\be
\label{eq:Lintelem}
\mcl{L}_{\inter} = \sum_{n = 0}^{4} \beta_{n} e_{n} (\gamma),
\ee
where $\beta_{n}$ are dimensionless coupling parameters and $e_{n}(\gamma)$ are the elementary symmetric polynomials of the trace of the so-called ``square-root matrix,'' defined by
\be
\label{eq:gamdef}
\tn{\gamma}{^\mu_\alpha} \tn{\gamma}{^\alpha_\nu} = \tn{g}{^\mu^\alpha} \tn{f}{_\alpha_\nu}.
\ee
The coupling strength of the interaction term is controlled by the product of an "effective" Planck mass
\be
M_{\mathrm{eff}}^2= {\lb \frac{1}{M_g^2}+\frac{1}{M_f^2} \rb}^{-1}.
\ee
and a continuous scaling parameter $m$ that is related (but not equal) to the mass of the graviton. The latter could be absorbed into the $\beta_n$ parameters, but we will not do so here so that the $\beta_n$ parameters are of order unity\footnote{If one wished to consider a $\beta_{n}$ that is larger than order unity, then one could simply increase the scaling parameter $m$.}. 

Without any loss of generality, the interaction term can be rewritten as
\be
\label{eq:Lintbtoc}
\mcl{L}_{\inter} = \sum_{n = 0}^{4} c_{n} e_{n} (K),
\ee
where $\tn{K}{^\mu_\nu} = \tn{\delta}{^\mu_\nu} - \tn{\gamma}{^\mu_\nu}$. This form of the interaction term allows us to easily enforce asymptotic flatness by requiring $c_0 = c_1 = 0$, and to reduce the theory to that of Pauli and Fierz~\cite{Pauli:1939xp} in the linearized limit of a massive spin 2 field by requiring $c_2 = 1$. The mapping from the remaining $c_{n}$ to $\beta_n$ is then
\be
\label{eq:btoc}
\beta_n = {\lb -1 \rb}^{n+1} \lsb \frac{1}{2} \lb 3 - n \rb \lb 4-n \rb - \lb 4 - n \rb c_3 - c_4 \rsb.
\ee
With this at hand, the interaction term becomes simply
\bal
\label{eq:Lint}
\mathcal{L}_{\inter} &= \frac{1}{2} \lb \tn{K}{^\mu_\mu} \rb^2 - \frac{1}{2} \tn{K}{^\nu_\mu} \tn{K}{^\mu_\nu} \nn \\
& + \frac{c_{3}}{3!} \tn{\epsilon}{_\mu_\nu_\rho_\sigma} \tn{\epsilon}{^\alpha^\beta^\gamma^\sigma} \tn{K}{^\mu_\alpha} \tn{K}{^\nu_\beta}\tn{K}{^\rho_\gamma} \nn \\
& + \frac{c_4}{4!} \tensor{\epsilon}{_\mu_\nu_\rho_\sigma} \tn{\epsilon}{^\alpha^\beta^\gamma^\delta} \tn{K}{^\mu_\alpha} \tn{K}{^\nu_\beta}\tn{K}{^\rho_\gamma}\tn{K}{^\sigma_\delta}\, ,
\end{align}
where $\epsilon^{\alpha \beta \gamma \delta}$ is the Levi-Civita tensor.

One can now obtain the field equations by varying the action with respect to both metrics. Doing so directly, however, is somewhat cumbersome due to the dependence of the interaction term on the square-root metric. Following~\cite{Volkov:2012wp}, one can instead introduce two tetrad fields, $\tn{e}{^\mu_A}$ and $\tn{\omega}{^B_\nu}$, defined via 
\be
\label{eq:gftetdef}
\tn{g}{^\mu^\nu} = \tn{\eta}{^A^B} \tn{e}{^\mu_A} \tn{e}{^\nu_B} \qquad\text{and}\qquad \tn{f}{_\mu_\nu} = \tn{\eta}{_A_B} \tn{\omega}{^A_\mu} \tn{\omega}{^B_\nu}.
\ee
By imposing the symmetry condition
\be
\label{eq:tetsymcond}
\tn{e}{^\mu_A} \tn{\omega}{_B_\mu} = \tn{e}{^\mu_B} \tn{\omega}{_A_\mu},
\ee
the square-root matrix can then be rewritten as
\be
\label{eq:gamtetdef}
\tn{\gamma}{^\mu_\nu} = \tn{e}{^\mu_A} \tn{\omega}{^A_\nu}.
\ee
With this at hand, one can then vary the action with respect to these tetrad fields to obtain the field equations 
\bal
\label{eq:FEg}
\tn{G}{^\mu_\nu} &= 8 \pi \tn{T}{^\mu_\nu} + m^2 \cos^2{\eta} \; \tn{V}{^\mu_\nu}, \\
\label{eq:FEf}
\tn{\mathcal{G}}{^\mu_\nu} &= m^2 \sin^2{\eta} \; \tn{\mathcal{V}}{^\mu_\nu}.
\end{align}
where we have introduced the new coupling constant ${\eta \in [0,\frac{\pi}{2}]}$ via~\cite{Volkov:2012wp} 
\be
\tan^2{\eta} = \frac{\mcl{G}}{G}.
\ee
The auxiliary reduced Planck mass is then simply $M_f^2 = {M_g^2} {\cot^2{\eta}}$, while the reduced effective Planck mass is $
M_{\mathrm{eff}}^2 = M_g^2 \cos^2{\eta}$. The Einstein tensors $\tn{G}{^\mu_\nu}$ and $\tn{\mathcal{G}}{^\mu_\nu}$ are those associated with the physical and auxiliary metrics respectively, while $\tn{T}{^\mu_\nu}$ is the matter stress-energy tensor, which again couples only to the physical metric. The interaction tensors, $\tn{V}{^\mu_\nu}$ and $\tn{\mathcal{V}}{^\mu_\nu}$, are defined by
\bal
\label{eq:Vgdef}
\tn{V}{^\mu_\nu} &= \tn{\tau}{^\mu_\nu} - \tn{\delta}{^\mu_\nu} \mathcal{L}_{\inter}, \\
\label{eq:Vfdef}
\tn{\mathcal{V}}{^\mu_\nu} &= - \frac{\sqrt{-g}}{\sqrt{-f}} \tn{\tau}{^\mu_\nu},
\end{align}
with
\bal
\label{eq:taudef}
\tn{\tau}{^\mu_\nu} &= -3 \, \tn{\gamma}{^\mu_\nu}+\tn{\gamma}{^\mu_\mu} \tn{\gamma}{^\nu_\nu} - \tn{\gamma}{^\mu_\alpha} \tn{\gamma}{^\alpha_\nu} \nn \\
& - \frac{c_{3}}{2} \tn{\epsilon}{_\nu_\rho_\lambda_\sigma} \tn{\epsilon}{^\alpha^\beta^\delta^\sigma} \tn{\gamma}{^\mu_\alpha} \tn{K}{^\rho_\beta}\tn{K}{^\lambda_\delta} \nn \\
& - \frac{c_{4}}{6} \tn{\epsilon}{_\nu_\rho_\lambda_\sigma} \tn{\epsilon}{^\alpha^\beta^\chi^\delta} \tn{\gamma}{^\mu_\alpha} \tn{K}{^\rho_\beta}\tn{K}{^\lambda_\chi} \tn{K}{^\sigma_\delta}\,.
\end{align}

The effective mass of the graviton $\mu$ is determined from a linear analysis of the field equations about flat spacetime~\cite{DeFelice:2013nba}
\be
\label{eq:gravitonmass}
\mu = m \; \left[\frac{{\lb 1+\cot^2{\eta}\, \xi_c^2 \rb^{1/2} \Gamma_c^{1/2}}}{\cot{\eta}\,\xi_c}\right],
\ee
where $\Gamma_{c} \equiv \beta_1 \xi_{c} + 4 \beta_2 \xi_{c}^2 + 6 \beta_3 \xi_{c}^3$ and $\xi_c$ is the critical value of the ratio of the cosmological scale factor associated with the auxiliary metric to that associated with the physical metric when the cosmological matter energy density asymptotes to zero. One can show that $\xi_c$ is a solution to the polynomial equation~\cite{DeFelice:2013nba}
\bal
\label{eq:xic}
\begin{split}
0 &= \frac{\beta_1}{\cot^2{\eta}\, \xi_c} + \lb \frac{6 \beta_2}{\cot^2{\eta}} - \beta_0 \rb + \lb \frac{18 \beta_3}{\cot^2{\eta}} - 3 \beta_1\rb \xi_c \\
 &+ \lb \frac{24 \beta_4}{\cot^2{\eta}} - 6 \beta_2 \rb \xi_c^2 - 6 \beta_3 \xi_c^3,
\end{split}
\end{align}
Thus, the effective graviton mass is a function of all 4 coupling parameters. We see that $\eta$ does not completely control the mass of the graviton, but rather this is really determined by all coupling parameters, and in particular the scaling mass $m$.

The field equations also lead to certain conservation relations. Taking the g- and f-metric covariant divergences of Eqs.~\eqref{eq:FEg} and~\eqref{eq:FEf}, we find
\begin{align}
8 \pi \; \; {}_{\g}\!\nabla^{\nu} \tn{T}{^\mu_\nu} &= - m^2 \cos^2{\eta} \; \; {}_{\g}\!\nabla^{\nu}\tn{V}{^\mu_\nu}, \\
{}_{\f}\!{\nabla}^{\nu} \tn{\mathcal{V}}{^\mu_\nu} &= 0\,,
\end{align}
where the Bianchi identities require that ${}_{\g}\!\nabla^{\nu} \tn{G}{^\mu_\nu}  = 0 = {}_{\f}\!\nabla^{\nu} \tn{\mathcal{G}}{^\mu_\nu}$. The diffeomorphism invariance of the action, however, guarantees that the matter stress-energy tensor is independently conserved~\cite{Volkov:2011an}
\begin{align}
\label{eq:consT}
{}_{\g}\!\nabla^{\nu} \tn{T}{^\mu_\nu} &= 0\,,
\end{align}
which then implies that
\begin{align}
\label{eq:consV}
{}_{\g}\!\nabla^{\nu}\tn{V}{^\mu_\nu} &= 0\,.
\end{align}
%

\section{Slowly-rotating Neutron stars in Massive Bigravity within the Hartle-Thorne Approximation}
\label{sec:nrBG}

In this section, we first introduce our choice of metric ansatzes, following the Hartle and Thorne approximation~\cite{1967ApJ...150.1005H,Hartle:1968si}, and then we describe our choice of stress-energy tensors to model neutron stars. We conclude this section with a derivation of the field equations in a slow-rotation expansion and a recasting of them into a form amenable to numerical integration. 

\subsection{Metric Tensors for Slowly-Rotating Spacetimes}

Following the approach of Hartle and Thorne~\cite{1967ApJ...150.1005H,Hartle:1968si}, we begin with the following metric ansatzes
\bal
\label{eq:gmet}
ds^2_{\mrm{g}} &= - e^{2 \nu} dt^2 + e^{2 \lambda} dr^2 + r^2 \lsb d\theta^2 + \ssqth \lb d\phi - \omega dt \rb^2 \rsb, \\
\label{eq:fmet}
ds^2_{\mrm{f}}&= - e^{2 \alpha} dt^2 + e^{2 \beta} dr^2 + U^2 \lsb d\theta^2 + \ssqth \lb d\phi - w dt \rb^2 \rsb,
\end{align}
where each metric function $\nu$, $\lambda$, $\alpha$, $\beta$, $U$, $\omega$, and $w$ are functions of $r$ and $\theta$ only. The above metric ansatzes are the result of an expansion in slow rotation to second order of the most general, stationary and axisymmetric metric. We make here a \emph{symmetric} choice of ansatzes: we choose the same functional form for the g- and the f-metrics. This is a choice because, although the g-metric must take the form above when imposing stationarity and axisymmetry, the f-metric could in principle take a different form.  We will see below that this symmetric choice is sufficient to find consistent solutions. 

\subsection{Neutron Star Stress-Energy Tensor}

We model the matter field as a perfect fluid that is uniformly rotating with angular velocity $\Omega$. The matter stress-energy tensor for such a fluid is
\be
\label{eq:perfltn}
\tn{T}{^{\mrm{(m)}}^\mu_\nu} = \lb p+\rho \rb \tn{u}{^\mu} \tn{u}{_\nu} + p \, \tn{g}{^\mu_\nu},
\ee
with density $\rho$, pressure $p$ and four-velocity
\be
\label{eq:4vel}
\tn{u}{^\mu} = \lsb \tn{u}{^t},0,0,\Omega \, \tn{u}{^t}\rsb.
\ee
The component $\tn{u}{^t}$ is obtained from the normalization condition $\tn{u}{^\mu} \tn{u}{_\mu} = -1$,
\be
\tn{u}{^t} = \frac{1}{e^{\nu}} + \frac{1}{2} \frac{r^2 \ssqth}{e^{3 \nu}} \lb \omega^2 - 2 \omega \Omega + \Omega^2 \rb + \mathcal{O} \lb \Omega^4 \rb.
\ee
Ignoring terms of $\mathcal{O} \lb \Omega^2 \rb$, the stress-energy tensor in matrix form is then
\be
\label{eq:Tm}
\tn{T}{^{\mrm{(m)}}^\mu_\nu} = \begin{pmatrix} -\rho &0&0& \frac{r^2 \ssqth}{e^{2 \nu}} \lb \Omega-\omega\rb \lb p+\rho \rb \\ 0 & p &0&0 \\ 0&0& p &0 \\ -\Omega\left(p+\rho\right) &0&0& p \end{pmatrix}.
\ee

In order to close our system of equations we require a relationship between the pressure and the density. We here only consider barotropic equations of state, namely $p = p(\rho)$, that are obtained numerically by solving certain systems of equations in nuclear physics. In particular, we focus on the Akmal-Pandharipande-Ravenhall (APR)~\cite{Akmal:1998cf}, Lattimer-Swesty (LS220)~\cite{LATTIMER1991331}, and Shen et al.~\cite{Shen:1998gq} equations of state. APR uses a variational chain summation method combined with a leading-order relativistic correction and three-nucleon interactions. LS220 uses a compressible liquid-drop model for nuclei and includes multiple additional nuclear interactions. Shen is constructed from relativistic mean field theory over a wide range of baryon mass densities, temperatures, and proton fractions. These equations of state allow for neutron stars with masses above the PSR J1614-2230 limit (above $1.97 M_{\odot}$~\cite{Demorest:2010bx}) in GR.

\subsection{Interaction Tensor}

Before we can find the modified field equations, we must first evaluate the interaction tensors, which in turn require the calculation of the square-root matrix. Let us then begin with the latter, using Eq.~\eqref{eq:gamdef} and an axisymmetric choice for the tetrad fields $e_{A}^{\mu}$ and $\omega^{A}_{\mu}$:
\be
  \begin{split}
    e_{t}^{\mu} &= \left[M,0,0,K \right],\\
    e_{r}^{\mu} &= \left[0,B,0,0 \right],\\
    e_{\theta}^{\mu} &= \left[0,0,C,0 \right], \\
    e_{\phi}^{\mu} &= \left[X,0,0,Y \right],
  \end{split}
\qquad 
  \begin{split}
    \omega^{t}_{\mu} &= \left[m,0,0,k \right],\\
    \omega^{r}_{\mu} &= \left[0,b,0,0 \right], \\
    \omega^{\theta}_{\mu} &= \left[0,0,c,0 \right], \\
    \omega^{\phi}_{\mu} &= \left[x,0,0,y \right].
  \end{split}
\ee
where $(M, K, B, C, X, Y, m, k, b, c, x,y)$ are arbitrary functions of $(r,\theta)$. Using Eqs.~\eqref{eq:gftetdef} and~\eqref{eq:gamtetdef}, we then find
\bal
\label{eq:ggentet}
\tn{g}{^\mu^\nu} &= \begin{pmatrix} - M^2 + Y^2 &0&0& -M \, K + X \, Y \\ 0 & B^2 &0&0 \\ 0&0& C^2 &0 \\-M \, K + X \, Y &0&0& -K^2 + X^2 \end{pmatrix}, \\
\label{eq:fgentet}
\tn{f}{_\mu_\nu} &= \begin{pmatrix}  -m^2 +y^2&0&0& -m \, k + x \, y \\ 0 & b^2 &0&0 \\ 0&0& c^2 &0 \\ -m \, k + x \, y &0&0& -k^2 + x^2 \end{pmatrix}, \\
\label{eq:gamtet}
\tn{\gamma}{^\mu_\nu} &= \begin{pmatrix} M m + Y y &0&0& M k + Y x \\ 0 & B b &0&0 \\ 0&0& C c &0\\ K m + X y &0&0& K k + X x\end{pmatrix}.
\end{align}
Comparing the above metric tensors to the metric ansatzes in Eqs.~\eqref{eq:gmet} and~\eqref{eq:fmet}, and comparing the square-root matrix above to that obtained from Eq.~\eqref{eq:gamdef} computed with the metric ansatzes of Eqs.~\eqref{eq:gmet} and~\eqref{eq:fmet}, the only solution that is spherically symmetric in the non-rotating limit~\cite{Volkov:2012wp} is
\be
  \begin{split}
    M &= {e^{-\nu}},\\
    X &= \left({r \sin{\theta}}\right)^{-1},\\
    m &= a, \\
    x &= U \sin{\theta},
  \end{split}
\qquad 
  \begin{split}
    K &= {\omega} \; {e^{-\nu}},\\
    Y &= 0, \\
    k &= 0, \\
    y &= - w U \sin{\theta}.
  \end{split}
\ee
With this solution, we can use the tetrad symmetry condition in Eq.~\eqref{eq:tetsymcond} to show that $\omega = w$, i.e.~frame dragging is identical in both spacetimes. With the basis solution found above, the square-root matrix is
\be
\label{eq:gamsol}
\tensor{\gamma}{^\mu_\nu} = \begin{pmatrix} {e^{\alpha-\nu}} &0&0& 0 \\ 0 & {e^{\beta-\lambda}} &0&0 \\ 0&0& {U}/{r} &0 \\ \omega \left( {e^{\alpha-\nu}} - {U}/{r} \right) &0&0& {U}/{r} \end{pmatrix}.
\ee

With the square-root matrix in hand, we can now calculate the full interaction tensors, $\tn{V}{^\mu_\nu}$ and $\tn{\mathcal{V}}{^\mu_\nu}$, from Eqs.~\eqref{eq:Vgdef} and~\eqref{eq:Vfdef}:
\allowdisplaybreaks[4]
\bal%
\label{eq:Vcomponents}
\begin{split}
\tensor{V}{^t_t} &= \kappa_1 \frac{e^{\beta}}{e^{\lambda}} + \kappa_2, \\
\tensor{V}{^r_r} &= \kappa_1 \frac{e^{\alpha}}{e^{\nu}} + \kappa_2, \\
\tensor{V}{^\theta_\theta} =\tensor{V}{^\phi_\phi} &= \frac{e^{\alpha}}{e^{\nu}} \left(\kappa_3 \frac{e^{\beta}}{e^{\lambda}} - \kappa_4 \right) - \kappa_4 \frac{e^{\beta}}{e^{\lambda}} + \kappa_5, \\
\tensor{V}{^\phi_t} &= \omega \left( \frac{U}{r} - \frac{e^{\alpha}}{e^{\nu}} \right) \left( \kappa_3 \frac{e^{\beta}}{e^{\lambda}} - \kappa_4 \right), \\
\tensor{\mathcal{V}}{^t_t} &= \kappa_6 \frac{e^{\lambda}}{e^{\beta}} + \kappa_7, \\
\tensor{\mathcal{V}}{^r_r} &= \kappa_6 \frac{e^{\nu}}{e^{\alpha}} + \kappa_7, \\
\tensor{\mathcal{V}}{^\theta_\theta} =\tensor{\mathcal{V}}{^\phi_\phi} &= \frac{e^{\nu}}{e^{\alpha}} \left( \kappa_8 \frac{e^{\lambda}}{e^{\beta}} + \kappa_9 \right) + \kappa_9 \frac{e^{\lambda}}{e^{\beta}} + \kappa_{10}, \\
\tensor{\mathcal{V}}{^\phi_t} &= \omega \left[ - \frac{e^{\nu}}{e^{\alpha}} \left( \kappa_8 \frac{e^{\lambda}}{e^{\beta}} + \kappa_9 \right) + \kappa_8 \frac{e^{\lambda}}{e^{\beta}} + \kappa_9 \frac{r}{U} \right], \\
\end{split}
\end{align}
where
\bal%
\label{eq:kappacomponents}
\begin{split}
\kappa_1 &= -3 + 3 c_3+c_4 - (4 c_3+2 c_4-2) \frac{U}{r} + (c_3+c_4) \frac{U^2}{r^2}, \\
\kappa_2 &= 6 - 4 c_3 - c_4 + (6 c_3+ 2 c_4 - 6) \frac{U}{r} - (2 c_3 +c_4 - 1)\frac{U^2}{r^2}, \\
\kappa_3 &= - (2 c_3+c_4+1) +(c_3+c_4)\frac{U}{r}, \\
\kappa_4 &= -(3 c_3+c_4+3) + (2 c_3+ c_4+1) \frac{U}{r}, \\
\kappa_5 &= -(4 c_3+c_4+6) + (3 c_3+c_4 + 3)\frac{U}{r}, \\
\kappa_6 &= c_3+c_4 - (4 c_3+2 c_4-2) \frac{r}{U}+(3 c_3+c_4-3)\frac{r^2}{U^2}, \\
\kappa_7 &= -c_4+(2 c_3+2 c_4)\frac{r}{U} - (2 c_3 +c_4 - 1)\frac{r^2}{U^2}, \\
\kappa_8 &= -(2 c_3+c_4+1) + (3 c_3+c_4 + 3)\frac{r}{U}, \\
\kappa_9 &= c_3+c_4- (2 c_3+c_4+1) \frac{r}{U}, \\
\kappa_{10} &= -c_4 + (c_3+c_4)\frac{r}{U}. \\
\end{split}
\end{align}

\subsection{Modified Field Equations}

Combining the above results we can write the equations of stellar structure for a neutron star in massive bigravity. The approach is similar to that in GR, so let us begin by deriving the field equations to zeroth-order in slow-rotation. Making the familiar GR substitution
\be
\label{eq:lamdef}
e^{2 \lambda} = \left({1-\frac{2 m_g}{r}}\right)^{-1},
\ee
where $m_{g}$ is a new function of $r$ only, the $(t,t)$ and $(r,r)$ component of the g-metric field equations become
\bal
\label{eq:dmgdr}
&\frac{d m_g}{d r} = 4 \pi r^2 \rho - \frac{1}{2} m^2 \cos^2{\eta} \; r^2 \tn{V}{^t_t}, \\
\label{eq:dnudr}
&\frac{d \nu}{d r} = \frac{4 \pi r^3 p + m_g}{r \lb r-2 m_g \rb} + \frac{1}{2} \frac{m^2 \cos^2{\eta} \; r^2 \tn{V}{^r_r}}{r-2 m_g}.
\end{align}
Notice that as $m \rarr 0$ or $\eta \rarr \eta/2$, $m_g(r)$ becomes the GR total enclosed mass of the neutron star inside radius $r$.

We can express the f-metric equations in a similar manner. Using the substitution
\be
\label{eq:betadef}
e^{2 \beta} = {U^{\prime}}^2 \left({1-\frac{2 m_f}{U}}\right)^{-1},
\ee
where $m_{f}$ is a new function of $r$ only and $U^{\prime} = {d U}/{d r}$, the $(t,t)$ and $(r,r)$ components of the f-metric field equations become
\bal
\label{eq:dmfdr}
&\frac{d m_f}{d r} = -\frac{1}{2} m^2 \sin^2{\eta} \; U^2 \; U^{\prime} \; \tn{\mathcal{V}}{^t_t}, \\
\label{eq:dalphadr}
&\frac{d \alpha}{d r} = U^{\prime} \lsb \frac{m_f}{U \lb U-2 m_f \rb} + \frac{1}{2} \frac{m^2 \sin^2{\eta} U^2 \tn{\mathcal{V}}{^r_r}}{U-2 m_f} \rsb.
\end{align}

We have so far obtained four equations for the four metric function unknowns $(\nu, m_g, \alpha, m_f)$, but we still need equations for $(p,\rho,U)$. The equation of state provides a relation between $p$ and $\rho$. Conservation of the matter stress-energy tensor, Eq.~\eqref{eq:consT}, leads to the equation
\be
\label{eq:dpdr}
\frac{d p}{d r} = -\lb p+\rho \rb \frac{d \nu}{d r}.
\ee
Conservation of the $\tn{V}{^\mu_\nu}$ tensor, Eq.~\eqref{eq:consV}, leads to the equation
\bal
\label{eq:Vgcons}
&\kappa_1 \frac{e^{\alpha}}{e^{\nu}} \frac{d \alpha}{d r} = \frac{r}{2} \lb \kappa_3 \frac{e^{\alpha}}{e^{\nu}}-\kappa_4 \rb U^{\prime} \nn \\
&+ \lsb \kappa_1 \frac{d \nu}{d r} - \frac{2}{r}\lb \kappa_3 \frac{e^{\alpha}}{e^{\nu}}-\kappa_4 \rb \rsb \sqrt{\frac{U \lb r-2 m_g \rb}{r \lb U-2 m_f \rb}} U^{\prime}\,,
\end{align}
which then closes our system of ordinary differential equations at zeroth-order in spin.

Let us now consider the modified field equations to first-order in slow-rotation. To this order, $\omega$ is the only free metric function, and thus, we only need a single differential equation. Investigating the $\lb t,\phi \rb$ component of the g-metric field equation, we find
\allowdisplaybreaks[4]
\begin{widetext}
\begin{align}
\label{eq:d2wdr2full}
&\frac{\pd^2 \omega}{\pd r^2} + \frac{4}{r} \lsb 1 - \frac{\pi r^3 \lb p+\rho \rb}{r - 2 m_g} - \frac{m^2 \cos^2{\eta}\, r^3}{8 \lb r-2 m_g \rb} \lb \tn{V}{^r_r} - \tn{V}{^t_t} \rb \rsb \frac{\pd \omega}{\pd r} 
\nn \\
&+ \frac{16 \pi r \lb p + \rho \rb}{r - 2 m_g} \lb \Omega - \omega \rb + \frac{1}{r \lb r-2 m_g \rb} \lb \frac{\pd^2 \omega}{\pd \theta^2} + 3 \frac{\cos{\theta}}{\sin{\theta}} \frac{\pd \omega}{\pd r} \rb \nn \\
& + \frac{1}{2} \frac{m^4 \cos^4{\eta}\, r^4 \tn{V}{^r_r} \lb \tn{V}{^t_t}-\tn{V}{^r_r} \rb}{\lb r-2 m_g \rb^2} \omega
- m^2 \cos^2{\eta} 
\nn \\
&\times \lsb \frac{r \lb \tn{V}{^r_r}-\tn{V}{^t_t} \rb }{\lb r-2 m_g \rb^2} \lb 4 \pi r^3 p + m_g \rb+ \frac{r^2}{r-2 m_g} \frac{d \tn{V}{^r_r}}{d r} + \frac{2 r}{r - 2 m_g} \lb \tn{V}{^r_r}+\tn{V}{^\phi_\phi} \rb \rsb \omega = 0.
\end{align}
\end{widetext}
Following~\cite{1967ApJ...150.1005H}, we use a vector spherical harmonic decomposition
\be
\omega (r,\theta) = \sum_{l=1}^\infty \omega_{l}(r) \lb -\frac{1}{\sin{\theta}} \frac{d P_l(\cos{\theta})}{d \theta} \rb,
\ee
where $P_l(\cos{\theta})$ are Legendre polynomials, to decouple this equation. The polar term becomes
\be
\frac{\pd^2 \omega}{\pd \theta^2} + 3 \frac{\cos{\theta}}{\sin{\theta}} \frac{\pd \omega}{\pd \theta} = - \lb l+2 \rb \lb l-1 \rb \omega.
\ee
By imposing regularity at the neutron star core and asymptotic flatness, all $\omega_l$ must vanish except for the $l=1$ term, as expected. Equation~\eqref{eq:d2wdr2full} then becomes
\begin{widetext}
\bal
\label{eq:d2wdr2}
&\frac{\pd^2 \omega}{\pd r^2} + \frac{4}{r} \lsb 1 - \frac{\pi r^3 \lb p+\rho \rb}{r - 2 m_g} - \frac{m^2 \cos^2{\eta}\, r^3}{8 \lb r-2 m_g \rb} \lb \tn{V}{^r_r} - \tn{V}{^t_t} \rb \rsb \frac{\pd \omega}{\pd r} + \frac{16 \pi r \lb p + \rho \rb}{r - 2 m_g} \lb \Omega - \omega \rb  + \frac{1}{2} \frac{m^4 \cos^2{\eta} \,r^4 \tn{V}{^r_r} \lb \tn{V}{^t_t}-\tn{V}{^r_r} \rb}{\lb r-2 m_g \rb^2} \omega 
\nn \\
&- m^2 \cos^2{\eta} \lsb \frac{r \lb \tn{V}{^r_r}-\tn{V}{^t_t} \rb }{\lb r-2 m_g \rb^2} \lb 4 \pi r^3 p + m_g \rb+ \frac{r^2}{r-2 m_g} \frac{d \tn{V}{^r_r}}{d r} + \frac{2 r}{r - 2 m_g} \lb \tn{V}{^r_r}+\tn{V}{^\phi_\phi} \rb \rsb \omega = 0.
\end{align}
\end{widetext}
%

\subsection{Recasting of Modified Field Equations}

The equations of stellar structure presented thus far are not ideal for numerical integration in their present form. We here follow the approach of~\cite{Volkov:2012wp} to recast the differential equations into a simpler form. First, we substitute Eq.~\eqref{eq:dalphadr} into~\eqref{eq:Vgcons}, causing each $U^{\prime}$ to cancel exactly from the latter. We then use Eq.~\eqref{eq:dnudr} to solve for the quantity $e^{(\alpha-\nu)}$ in the resulting expression. The right-hand side of this expression is lengthy and will be denoted as the function $F$ that depends on the metric functions $m_g$, $m_f$, $r$, and $U$ and the coupling parameters $m$, $c_3$, $c_4$, and $\eta$: 
\be
\label{eq:Fdef}
\frac{e^{\alpha}}{e^{\nu}} = F \lsb r,m_g,m_f,U,m,c_3,c_4,\eta \rsb
\ee
This expression allows us to simplify our differential system because $\alpha$ and $\nu$ only appear in this particular combination. Thus, we insert Eq.~\eqref{eq:Fdef} into Eq.~\eqref{eq:dnudr} and~\eqref{eq:dalphadr}, to eliminate $\nu$ and $\alpha$ from our differential system.

To further simplify the equations, we insert Eqs.~\eqref{eq:lamdef} and~\eqref{eq:betadef} into the $\tn{V}{^t_t}$ term of Eq.~\eqref{eq:dmgdr} (and similarly in the $\tn{\mcl{V}}{^t_t}$ term of Eq.~\eqref{eq:dmfdr}). This separates the equations into two terms: one that is proportional to $U^{\prime}$ and one that is independent of $U^{\prime}$. Each of these terms are lengthy and will be denoted as separate functions $F_k = F_k \lsb r,m_g,m_f,U,m,c_3,c_4,\eta \rsb$ with $k \in (1,6)$; such that our simplified differential system is
\bal
\label{eq:dmgdrF}
\frac{d m_g}{d r} &= F_1 \frac{d U}{d r} + F_2 \\
\label{eq:dmfdrF}
\frac{d m_f}{d r} &= F_3 \frac{d U}{d r} + F_4 \\
\label{eq:dnudrF}
\frac{d \nu}{d r} &= F_5 \\
\label{eq:dalphadrF}
\frac{d \alpha}{d r} &= F_6 \frac{d U}{d r}.
\end{align}
A few of the $F_k$ functions are given explicitly in Appendix~\ref{app:Feq} to show an example of their structure.
 
To close the system, we now need an equation for the evolution of $U$. Taking a total derivative of both sides of Eq.~\eqref{eq:Fdef} and substituting Eqs.~\eqref{eq:dmgdrF} \textemdash~\eqref{eq:dalphadrF}, one can solve for ${d U}/{d r}$ to find
\be
\label{eq:DU}
\frac{d U}{d r} = \frac{\frac{\pd F}{\pd r} + \frac{\pd F}{\pd m_g} F_2 + \frac{\pd F}{\pd m_f} F_4 + F\,F_5}{F\,F_6 - \frac{\pd F}{\pd U} - \frac{\pd F}{\pd m_g} F_1 - \frac{\pd F}{\pd m_f} F_3},
\ee
provided the denominator does not vanish. The numerical problem then reduces to simultaneously solving Eqs.~\eqref{eq:dmgdrF},~\eqref{eq:dmfdrF}, and~\eqref{eq:DU}. Once this is done, one can insert $(m_{g}, m_{f},U)$ in the right-hand sides of Eqs.~\eqref{eq:dnudrF} and~\eqref{eq:dalphadrF} to solve for $(\alpha,\nu)$. Alternatively, once we have solved for $\alpha$ (or $\nu$), we can use Eq.~\eqref{eq:Fdef} to find $\nu$ (or $\alpha$).

\section{Numerical Construction of Neutron Stars in Massive Bigravity}
\label{sec:rotBG}

In this section, we describe how to numerically evolve the differential system obtained in the previous section, beginning with the system at zeroth-order in rotation and continuing with the first-order in rotation system. In general, we will use a shooting method to solve the equations: 
\begin{enumerate}
\item[(i)] choose boundary conditions at a small radius (close enough to the neutron star core) and at a very large radius (close enough to spatial infinity), 
\item[(ii)] numerically integrate from the small radius out to the surface and from the large radius in to the surface, \
\item[(iii)] check whether the solutions are continuous and differentiable at the surface, and 
\item[(iv)] if they are, declare this our solution; if they are not, choose different boundary conditions and repeat. 
\end{enumerate}
This shooting method then requires that we specify boundary conditions both at the core and at a large radius, which we do through a local analysis about $r = 0$ and $r= \infty$, as we will describe below. The key point of this method is that the boundary conditions will depend on integration constants that must be iterated over (or ``shot for'') to ensure continuity and differentiability.

\subsection{Zeroth-Order in Rotation}

\subsubsection{Local analysis at $r = 0$ and $r=\infty$}

At $r=0$, we Taylor expand $m_g$, $m_f$, $\nu$, $\alpha$, $p$, $\rho$, and $U$, insert the Taylor expansions into the differential system of the previous section, and equate coefficients of equal power in $r$ to find
\allowdisplaybreaks[1]
\begin{widetext}
\bal
\label{eq:coresol}
\begin{split}
U_{r \rarr 0} &= u r + \mathcal{O}(r^3), \\
m_{g_{r \rarr 0}} &= \frac{4}{3} \pi r^3 \rho_0 + \frac{1}{3} m^2 \cos^2{\eta}\, r^3 \lb u-1 \rb \lsb c_3 u^2 + \lb -\frac{7}{2} c_3 - \frac{3}{2} \rb u + \frac{5}{2} c_3 + 3 \rsb + \mathcal{O}(r^4), \\
m_{f_{r \rarr 0}} &= -\frac{1}{6} m^2 \sin^2{\eta}\, r^3 \lb u-1 \rb \lsb c_3 u^2 - 5 c_3 u+4 c_3 + 3 \rsb + \mathcal{O}(r^4), \\
\nu_{r \rarr 0} &= \nu_0 + \frac{1}{2} r^2 \lb 4 \pi p_0+\frac{4}{3} \pi \rho_0 \rb \nn \\
& + \frac{1}{2} m^2 \cos^2{\eta}\, r^2 \lcb \frac{1}{3} c_3 u^3 - \lb 2 c_3  + \frac{3}{2} \rb u + \frac{5}{3} c_3 + 2 + \lsb -c_3 u^2 + \lb 3 c_3 + 1 \rb u - 2 c_3 - \frac{3}{2} \rsb \frac{e^{\alpha_0}}{e^{\nu_0}} \rcb  + \mathcal{O}(r^3), \\
\alpha_{r \rarr 0} &= \alpha_0 + \frac{1}{6 u} m^2 \sin^2{\eta}\, r^2 \lcb c_3 u^3 - 3 c_3 u^2 + 2 c_3 + \frac{3}{2} -3 u \lsb c_3 u^2 + \lb -3 c_3 - 1 \rb u + 2 c_3 + \frac{3}{2} \rsb \frac{e^{\nu_0}}{e^{\alpha_0}}\rcb \mathcal{O}(r^3), \\
\rho_{r \rarr 0} &= \rho_0 + \mathcal{O}(r^2), \\
p_{r \rarr 0} &= p_0 + \mathcal{O}(r^2).
\end{split}
\end{align}
\end{widetext}
The quantity $\rho_0$ is the central density, i.e.~the density at the neutron star core, and it is a free parameter that effectively determines the mass of the star. Given $\rho_{0}$, the quantity $p_0$ is obtained from the equation of state, $p_0 = p[\rho_0]$. The quantities $u$ and $\nu_0$ are constants that must be shot for in order to guarantee continuity and differentiability of the solutions at the stellar surface. Given all of this, the quantity $\alpha_0$ is obtained from Eq.~\eqref{eq:Fdef}.

At $r=\infty$, we take a slightly different approach. If we insisted on a regular Taylor expansion of the metric functions, we would find that this ansatz does not satisfy the field equations. Instead, the field equations require a series expansion with a non-trivial controlling factor. This can be proved by parameterizing the local  behavior of the metric functions via 
\bal
\begin{split}
U_{r \rarr \infty} &= r + \delta \; U_\infty, \\
m_{g_{r \rarr \infty}} &= \delta \; m_{g_\infty}, \\
m_{f_{r \rarr \infty}} &= \delta \; m_{f_\infty}, \\
\nu_{r \rarr \infty} &= \delta \; \nu_\infty, \\
\alpha_{r \rarr \infty} &= \delta \;  \alpha_\infty, \\
\end{split}
\end{align}
which ensures asymptotic flatness, where $U_\infty$, $m_{g_\infty}$, $m_{f_\infty}$, $\nu_\infty$, and $\alpha_\infty$ are functions of $r$ and $\delta \ll 1$. Plugging the above into the field equations and keeping only terms of $\mathcal{O}(\delta)$ gives
\bal
\begin{split}
\label{eq:BCinf}
\frac{d}{d r} U_\infty &= \frac{m_{g_\infty} - m_{f_\infty}}{r}+2 \frac{m_g - m_f}{m^2 r^3}, \\
\frac{d}{d r} m_{g_\infty} &= m^2 \cos^2{\eta}\, r \, U_\infty + \cos^2{\eta} \frac{m_{g_\infty} - m_{f_\infty}}{r}, \\
\frac{d}{d r} m_{f_\infty} &= -m^2 \sin^2{\eta}\, r \, U_\infty - \sin^2{\eta}\frac{m_{g_\infty} - m_{f_\infty}}{r}, \\
\frac{d}{d r} \nu_{\infty} &= \frac{m_{g_\infty}}{r^2} + \cos^2{\eta}\frac{m_{g_\infty} - m_{f_\infty}}{r^2}, \\
\frac{d}{d r} \alpha_{\infty} &= \frac{2 m_{f_\infty}- m_{g_\infty}}{r^2} + \cos^2{\eta}\frac{m_{g_\infty} - m_{f_\infty}}{r^2}. \\
\end{split}
\end{align}
Solving this system of equations gives the asymptotic behavior,
\bal
\label{eq:infsol}
\begin{split}
U_{r \rarr \infty} &= r + \frac{m^2 r^2+m r +1}{m^2 r^2} B e^{-m r}, \\
m_{g_{r \rarr \infty}} &= A \sin^2{\eta} - B \cos^2{\eta} \lb m r+1 \rb e^{-m r}, \\
m_{f_{r \rarr \infty}} &=A \sin^2{\eta} + B \sin^2{\eta} \lb mr + 1 \rb e^{-m r}, \\
\nu_{r \rarr \infty} &= -\frac{A \sin^2{\eta}}{r} + \frac{2 B \cos^2{\eta}}{r} e^{-m r}, \\
\alpha_{r \rarr \infty} &= -\frac{A \sin^2{\eta}}{r} - \frac{2 B \sin^2{\eta}}{r} e^{-m r}, \\
\end{split}
\end{align}
where $A$ and $B$ are constants that are to be shot for by requiring continuity and differentiability of the solution at the stellar surface.

\subsubsection{Numerical Solutions}

Combining the above results, we can then solve for non-rotating neutron star solutions. The procedure is as follows. First, we specify a central density and choose an initial guess of the 4 matching parameters $(u,\nu_0,A,B)$. Second, using the boundary conditions at the neutron star core, we integrate outwards until the pressure vanishes, which denotes the surface of the star. Third, using the boundary conditions at infinity, we integrate inwards until the neutron star surface is reached. All integrations are carried out with an adaptive Runge-Kutta-Fehlberg algorithm with an error tolerance of $10^{-4}$.  Finally, we compare the metric functions at the surface to determine whether the solutions are continuous and differentiable also to a tolerance of $10^{-4}$. If they are not, we use a four-dimensional Newton-Raphson method to estimate a new guess for the 4 matching parameters, until the solutions are continuous and differentiable at the surface to the tolerance required.

\begin{figure}[htbp]
\begin{center}
\resizebox{8.75cm}{!}{\include{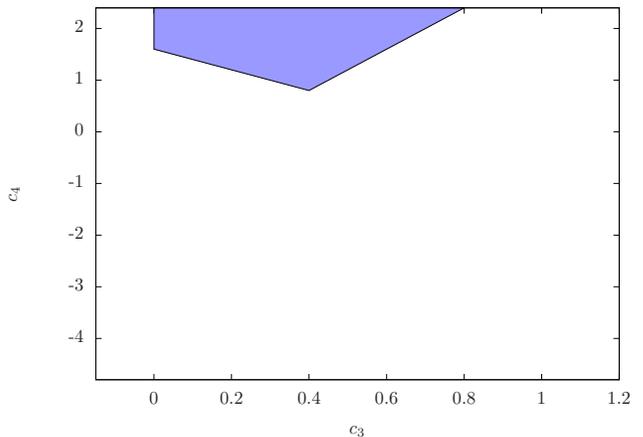}}
\caption{\label{fig:c3c4range}~[Color Online]. Numerical region of the expolored $(c_3,c_4)$ space. The figure is bounded by an estimate of the constrained region determined in~\cite{Cheung:2016yqr}. The shaded region shows the subregion in which we were able to numerically produce neutron stars solutions, with $m=\num{1.0e-7}\mrm{cm}^{-1}$ and $\eta=\pi/4$.}
\end{center}
\end{figure}

When solving for non-rotating neutron stars numerically, we must choose the region in coupling parameter space ($m,\eta,c_{3},c_{4})$ that we wish to explore. The range of $m$ was chosen to get a clear sense of the effect of bigravity modifications to GR. The range of $\eta$ chosen is its entire domain $(0,\pi/2$). The range of $(c_3,c_4)$ explored is approximately that of~\cite{Cheung:2016yqr}, which was obtained from analytic considerations of scattering amplitudes in massive gravity. However, we were unable to numerically produce stars in this entire $(c_3,c_4)$ subregion, only finding solutions in the shaded region shown in Fig.~\ref{fig:c3c4range}, which grows in size slightly for larger values of $m$. The most common cause of numerical issues is the denominator of Eq.~\eqref{eq:DU} becoming very small at a rate faster than the numerator. Whether this is due to a numerical limitation of our methods or whether it points to a deeper theoretical issue in massive bigravity is currently unclear and requires further study.

\begin{figure*}[htbp]
\begin{center}
\resizebox{8.75cm}{!}{\include{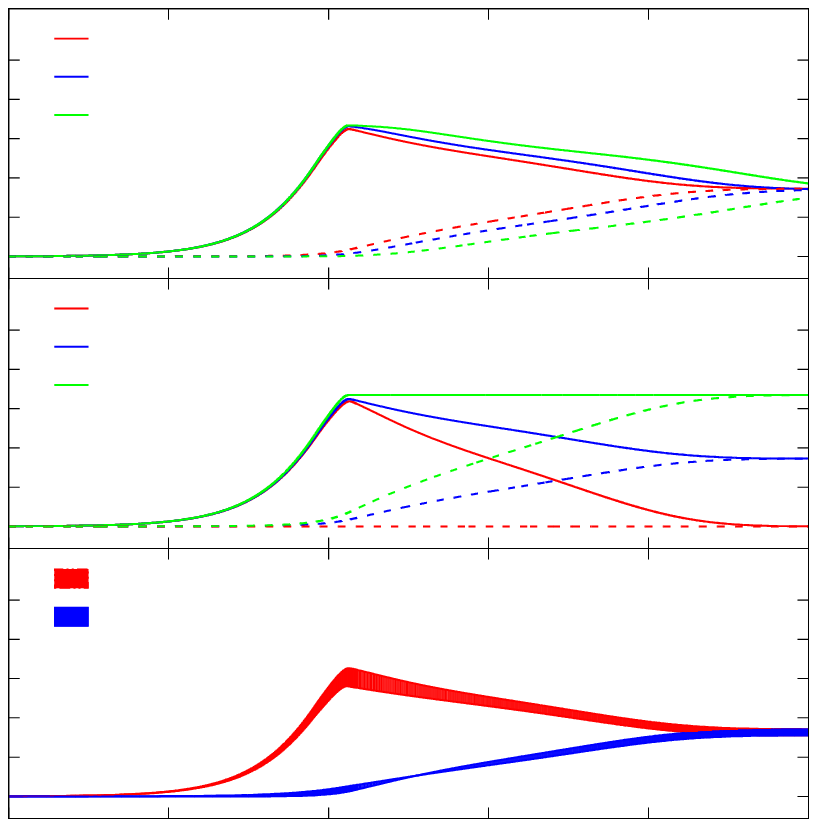}} \quad
\resizebox{8.75cm}{!}{\include{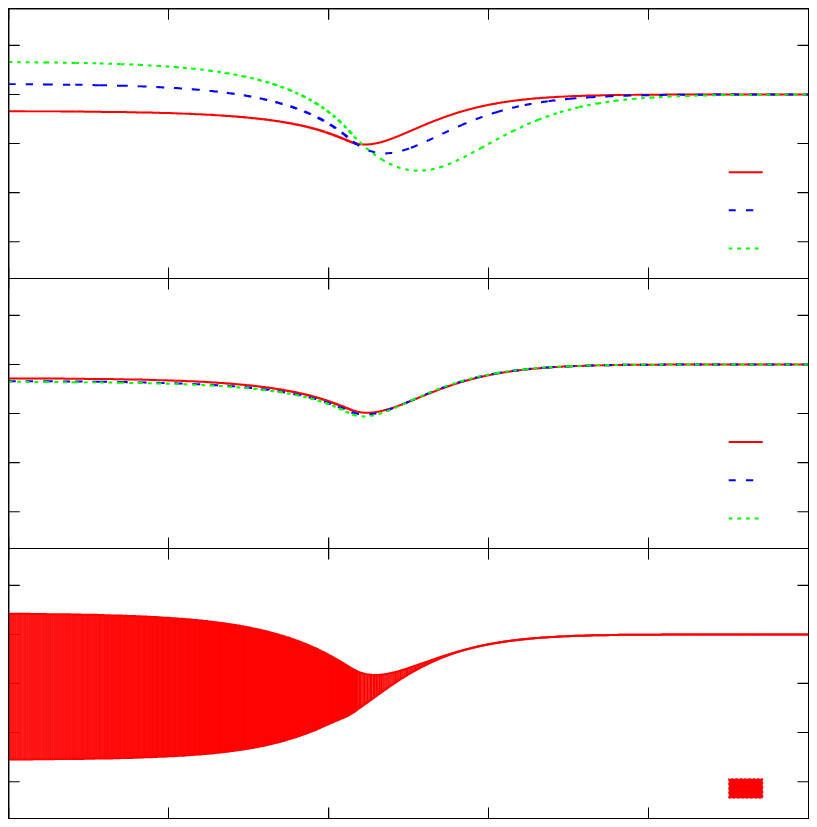}} \\
\resizebox{8.75cm}{!}{\include{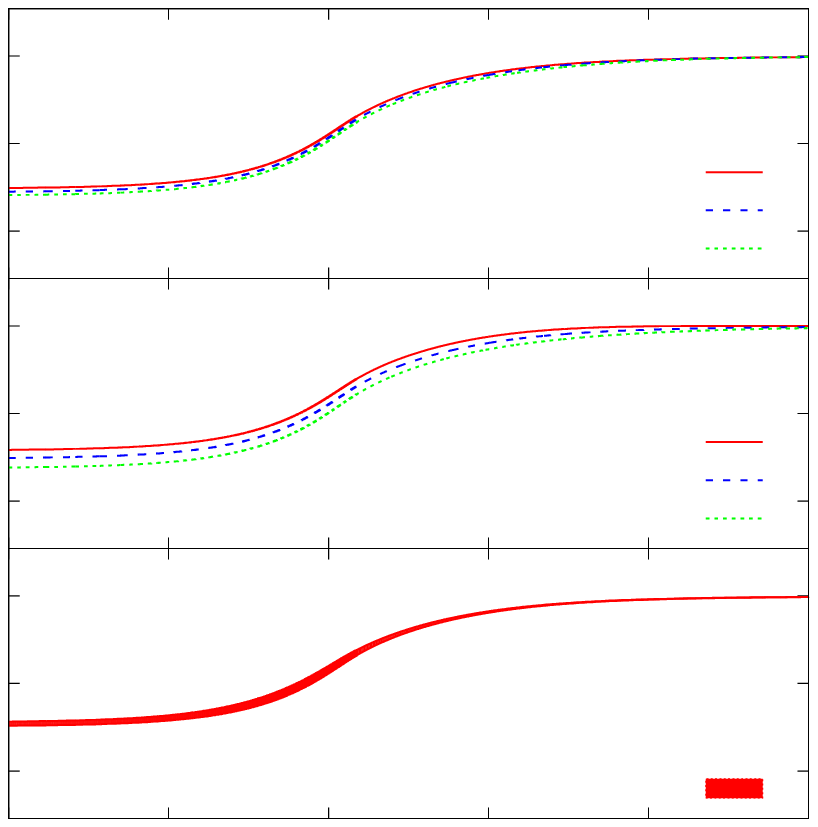}} \quad
\resizebox{8.75cm}{!}{\include{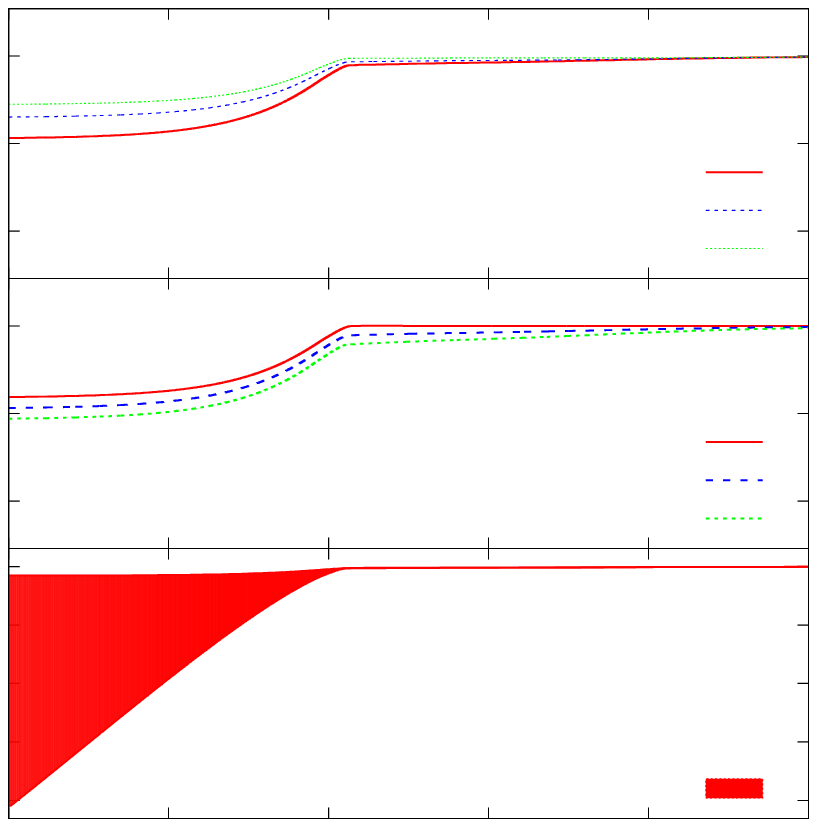}}
\caption{\label{fig:MgMf}~[Color Online]. Numerical solutions at zeroth-order in rotation as a function of radius for different choices of coupling parameters. In each subfigure, the parameter denoted in the caption is varied while all remaining parameters are set to a fixed value ($c_3 = 0.48$, $c_4 = 1.71$, $\eta = \pi/4$, $m = \num{3.0e-7}\mrm{cm}^{-1}$ where applicable). In the top two subfigures, we vary $m$ and $\eta$ for fixed $(c_{3},c_{4})$ through different color lines, while in the bottom subfigure we vary $c_3$ and $c_4$ for fixed $(m,\eta)$ through red or blue shaded regions. 
Top Left: Primary, $m_g$, and auxiliary, $m_f$, metric mass functions versus $r$.  Outside the star, the primary metric decouples from matter and the mass decreases to some asymptotic value unlike in GR. The parameter $\eta$ has the largest effect on this asymptotic value, while $m$ only alters the distance away from the stellar surface where both masses become equal, in accordance to the Vainshtein radius. Both $c_3$ and $c_4$ only broaden each mass profile curve, and so, they have the smallest effect on the asymptotic mass and radius observables.
Top Right: Ratio of the auxiliary radial function, $U$, to the primary radial function, $r$, versus $r$. The parameter $\eta$ has the smallest effect on $U$ overall, while $c_3$ and $c_4$ only affect it inside the star, but this effect is smaller than matter interactions. Thus, the parameter $m$ has the largest effect on $U$ outside the star.
Bottom Left: The primary $(t-t)$ metric function, $\nu$, versus $r$. In contrast to the $m_g$ and $m_f$ metric functions, the deviation of $\nu$ from GR only appears small because of the ${1}/{r}$ suppression of $\nu$ and $\alpha$ in Eq.~\eqref{eq:lamdef}.
Bottom Right: The auxiliary $(t-t)$ metric function, $\alpha$, versus $r$. Although the variation of $\alpha$ with $c_3$ and $c_4$ inside the star is large, this variation quickly approaches its asymptotic value at the surface of the star. Thus, the effect of $\alpha$ is subdominant to matter interactions in the region where it is present and it disappears in the region outside the star where the bimetric interactions dominate.
} 
\end{center}
\end{figure*}
\begin{figure*}[htbp]
\begin{center}
\resizebox{8.75cm}{!}{\include{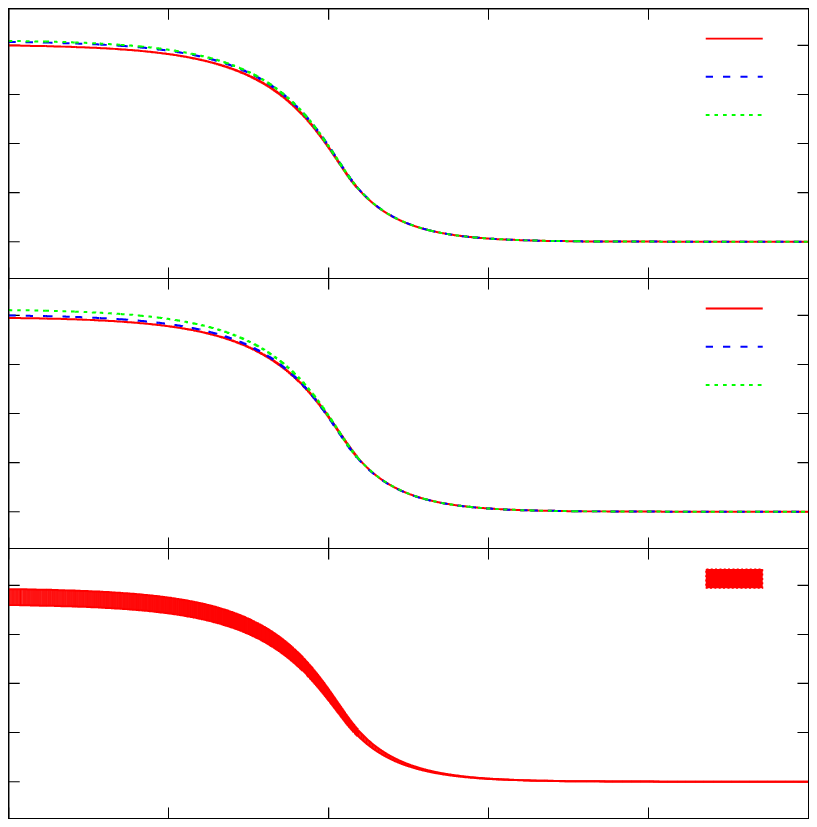}} \quad
\resizebox{8.75cm}{!}{\include{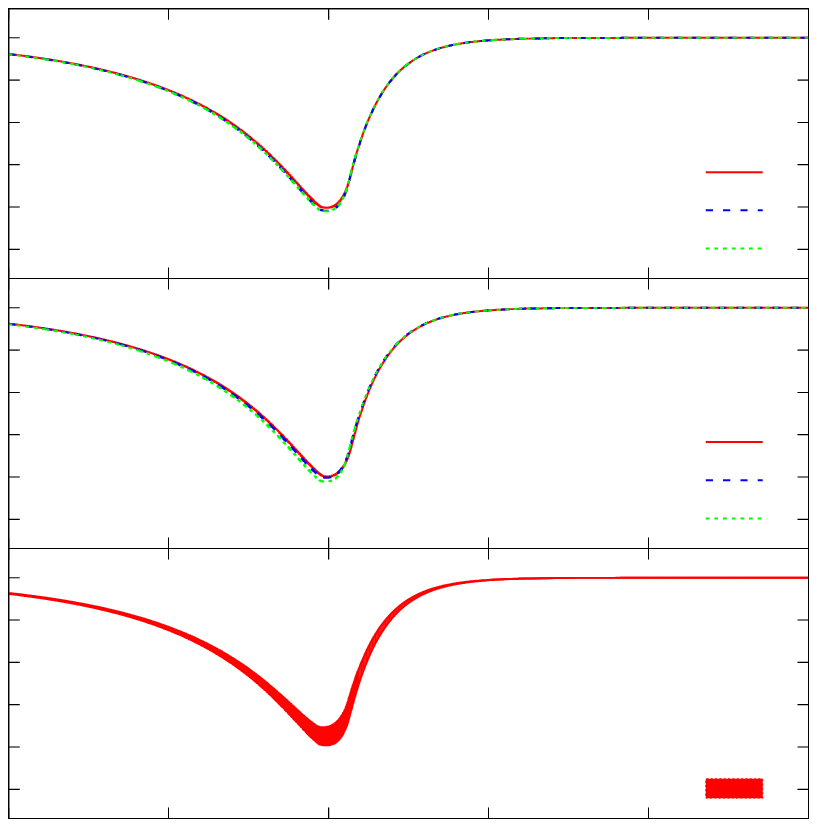}}
\caption{\label{fig:wdwdr}~[Color Online]. The first-order in rotation functions for both metrics, $\omega \, R$ (left) and ${d \omega}/{d r} R^2$ (right), versus radius. The format of this figure is identical to that of Fig.~\ref{fig:MgMf}. All four parameters seem to have a minimal effect on the first-order metric functions, unlike in the zeroth-order in rotation case. Observe that since $\omega = w$ from the symmetry condition in Eq.~\eqref{eq:tetsymcond} this is the first-order solution for both metrics.}
\end{center}
\end{figure*}

We present the non-rotating solutions for various couplings in Fig.~\ref{fig:MgMf}. All five metric profiles $m_g$, $m_f$, $U$, $\nu$, and $\alpha$ are shown as functions of radius for different coupling parameters ($m,\eta,c_3,c_4)$. The top left panel of Fig.~\ref{fig:MgMf} shows both the physical and the auxiliary mass functions versus radius. Observe that both increase monotonically from the core to the surface, with $m_g$ increasing much more rapidly than $m_f$ due to its direct matter coupling. Outside the star, the density term in Eq.~\eqref{eq:dmgdr} vanishes and the physical (auxiliary) mass monotonically decreases (increases) until it reaches its asymptotic value $m_g = A \sin^2{\eta}$, determined by Eq.~\eqref{eq:infsol}. Thus, the observable mass sufficiently far away from the star is determined by the matching parameter $A$, and the ratio of gravitational couplings $\tan^2{\eta} = {\mcl{G}}/{G}$. The parameter $\eta$ controls this asymptotic value, while the primary effect of $m$ is to determine how far from the surface both mass functions become effectively identical, i.e.~it controls the Vainshtein radius. Both $c_3$ and $c_4$ appear to broaden each profile rather than influence the asymptotic observable.

This panel also shows some interesting limiting behavior of the mass functions as one varies the coupling parameters. As $\eta \rarr \pi/2$, the g-metric in Eq.~\eqref{eq:FEg} decouples from the auxiliary metric and one recovers the Einstein field equation in GR. Therefore, in this limit, $m_g$ approaches the ADM mass at spatial infinity, $A = M_{\mathrm{ADM}}$. Moreover, in this limit the auxiliary gravitational constant $\mcl{G} \rarr \infty$ and the auxiliary Planck mass $M_f^2 \rarr 0$, recovering GR with an arbitrary decoupled auxiliary metric. On the other hand, in the limit as $\eta \rarr 0$, the auxiliary gravitational constant $\mcl{G} \rarr 0$ and the Plank mass $M_f^2 \rarr \infty$. In this limit, the auxiliary metric becomes flat from Eq.~\eqref{eq:FEf} and one recovers dRGT massive gravity.

The top right panel of Fig.~\ref{fig:MgMf} shows the auxiliary radial function, $U$, normalized to the primary radial function, $r$, as a function of radius. In this case, $\eta$ has a very small effect, while $c_3$ and $c_4$ have a very large impact inside the star. The parameter $m$ has a noticeable influence both inside and outside the star. Since the interactions with the auxiliary metric dominate outside the star where matter vanishes, the parameter $m$ dominates the behavior of $U$ outside the star.

The bottom left panel of Fig.~\ref{fig:MgMf} shows the $(t-t)$ component function of the primary metric, $\nu$, as a function of radius. The bigravity modifications to $\nu$ are much less pronounced than the modifications to the mass $m_g$. This is because the definition of $m_g$ in Eq.~\eqref{eq:lamdef} introduces a $1/r$ suppression in the metric function itself ($\lambda$ in this case). Were we to compare the original $(r-r)$ metric function $\lambda$ (instead of $m_{g}$) to the $(t-t)$ component $\nu$, the modification would be similarly less pronounced. 

The bottom right panel of Fig.~\ref{fig:MgMf} shows the auxiliary $(t-t)$ component function, $\alpha$, as a function of radius. In this case, $c_3$ and $c_4$ have a dramatic effect inside the star but this is almost entirely eliminated beyond the surface. This magnified effect inside the star is also evidenced by the sharp jump to its asymptotic value at the surface, shown in the top two portions of this panel. Once the stellar surface is reached, $\alpha$ immediately reaches its expected asymptotic value of $0$. This is peculiar because outside the star, absent of any dominant matter effects, the evolution of the system is driven by the interactions with the auxiliary metric. If $\alpha$ does not vary outside the star, then it cannot strongly influence these interactions.

\subsection{First-Order in Rotation}

\subsubsection{Local Analysis at $r = 0$ and $r = \infty$}

The numerical approach at first-order in spin is identical to that used at zeroth-order. The metric function we solve for is the first-order frame-dragging function $\omega$. After a Taylor expansion of Eq.~\eqref{eq:d2wdr2} about $r=0$ and equating coefficients in powers of $r$, the behavior of $\omega$ at the core of the star is identical to that in GR,
\be
\omega_{r \rarr 0} = \omega_0 - \frac{8 \pi}{5} r^2 \lb \rho_0 + p_0 \rb \lb \Omega - \omega_0 \rb + \mathcal{O}(r^3).
\ee
Just like $\rho_{0}$ determines the mass of the star, the constant $\Omega$ determines how fast the star is rotating and must be specified before integration. The constant $\omega_0$ must be shot for by requiring the solution to be continuous and differentiable at the stellar surface.

At $r= \infty$ we again cannot choose a simple Taylor expansion. Instead, let us propose a solution of the form
\be
\omega_{r \rarr \infty} = \omega_\gr+ \delta \; \omega_{\infty},
\ee
where again $\delta \ll 1$. The resulting equation to $\mathcal{O}(\delta^0)$ is the same as in GR
\be
\frac{d^2 \omega_\gr}{d r^2} + \frac{4}{r} \frac{d \omega_\gr}{d r} = 0,
\ee
whose solution is
\be
\label{eq:omegaGR}
\omega_\gr = \Omega - \frac{2 J}{r^3},
\ee
after choosing a proper integration constant and enforcing asymptotic flatness, where $J$ is the spin angular momentum of the rotating star; this must also be shot for by requiring the solution to be continuous and differentiable at the stellar surface. The equation at $\mathcal{O}(\delta)$ is
\be
\frac{d^2 \omega_{\infty}}{d r^2} + \frac{4}{r} \frac{d \omega_{\infty}}{d r} + \frac{J B \cos^2{\eta}}{e^{m r}} \lb -\frac{2 m^3}{r^3} +\frac{m^2}{r^4} -\frac{3 m}{r^5} - \frac{3}{r^6} \rb.
\ee
Solving the above equation gives
\bal
\begin{split}
\omega_{\infty} &= J B \cos^2{\eta} \frac{23 m^4}{12} \mrm{Ei}(m r) \\
& + \frac{J B \cos^2{\eta}}{e^{m r}} \lb -\frac{23 m^3}{12 r} + \frac{23 m^2}{12 r^2} + \frac{m}{6 r^3} + \frac{3}{2 r^4} \rb,
\end{split}
\end{align}
where $\mrm{Ei}(mr)$ is the exponential integral function. After expanding the exponential integral function about $r = \infty$ and combining the result with Eq.~\eqref{eq:omegaGR}, the behavior of $\omega$ as $r \rightarrow \infty$ is
\bal
\begin{split}
\omega_{r \rarr \infty} &= \Omega - \frac{2 J}{r^3} \\
& + \frac{J B \cos^2{\eta}}{e^{m r}} \lsb \frac{4 m}{r^3} - \frac{10}{r^4} + \frac{46}{m r^5}+ \mathcal{O}\lb \frac{1}{r^6} \rb \rsb.
\end{split}
\end{align}
%

\subsubsection{Numerical Solution}

We can now use these boundary conditions on $\omega$ to obtain slowly rotating neutron star solutions. The procedure is identical to that in the non-rotating case. We make an initial guess for $\omega_0$ and $J$ and integrate from the core and from spatial infinity to the surface of the star. Then, we use a two-dimensional Newton-Raphson method to estimate a new guess for $\omega_0$ and $J$ until we find a solution that is continuous and differentiable at the surface to the same tolerance as in the zeroth-order in rotation case. An example is presented in Fig.~\ref{fig:wdwdr} where we show the first-order frame dragging function $\omega$ and its derivative. As before, we independently vary $m$, $\eta$, and $c_3$ and $c_4$ but we find no noticeable trends from each of these parameters effects.

\begin{figure}[htbp]
\begin{center}
\resizebox{8.75cm}{!}{\include{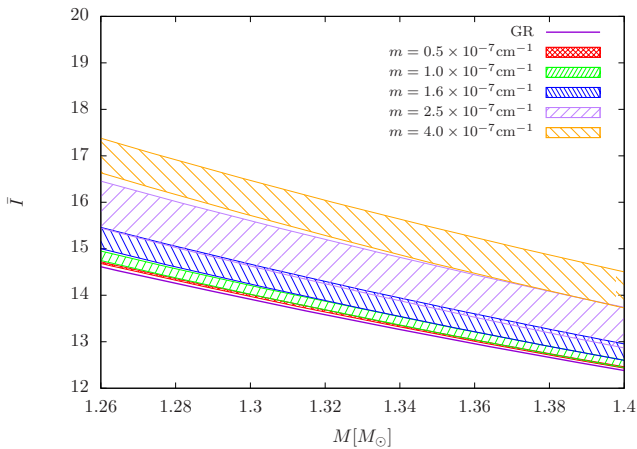}}
\caption{\label{fig:IMbar}~[Color Online]. Moment of inertia-mass relation with the APR equation of state for various $c_3$ and $c_4$. Each shaded region represent families of curves for a fixed value of $m$ and $\eta = {\pi}/{4}$. Although the broadening introduced by these couplings scales with $m$, they remain sub-dominant to variation in $\eta$.}
\end{center}
\end{figure}

Once the rotating star is constructed, the moment of inertia can be obtained from the definition, 
\be
I \equiv \frac{J}{\Omega}.
\ee
For convenience, we also define the dimensionless moment of inertia $\bar{I} \equiv {I}/{M^3}$, where $M$ is the mass function of the g-metric evaluated at the stellar surface. Figure~\ref{fig:IMbar} shows this dimensionless moment of inertia as a function of mass with the APR equation of state for various $(c_{3},c_{4})$ and $m$ values. Observe that the dimensionless moment of inertia is mostly controlled by $m$, with $(c_{3},c_{4})$ introducing a broadening in the curve; the larger $m$, the larger the broadening effect.

\section{Conclusions and Future Directions}
\label{sec:conc}

We have constructed slowly-rotating neutron stars in massive bigravity with various realistic equations of state and a range of coupling parameters. From these solutions, we have explored the mass-radius and the moment-of-inertia-mass relations in some detail. We have found that the bigravity coupling parameters $m$ and $\eta$ have the strongest impact on these relations, with significant deviations from GR arising when $m > 10^{-7}\mrm{cm}^{-1}$ and $\eta < \pi/2$. The results obtained here can thus serve as the foundations for an experimental study of bigravity in light of binary pulsar observations and low-mass X-ray binary observations. 

The fitting of a timing model to the observation of a sequence of radio pulses from binary pulsars allows one to extract a measure of certain post-Keplerian parameters~\cite{Ruggiero:2005dg}, like the rate of change of the orbital period and the Shapiro time delay parameter. These measurements can then be combined to infer the masses of the binary stars up to a given observational error. For example, for PSR J0737-3039A, fits to the timing model suggest that the primary pulsar PSR J0737-3039A has a mass of $(1.337 \pm 0.0037) M_{\odot}$~\cite{Lyne1153}. In a modified gravity theory, the post-Keplerian parameters are not only a function of the masses of the binary components, but also of other parameters in the theory. Thus, the measurement of more than two post-Keplerian parameters yields a test of GR~\cite{Damour:2007uf}. 

In bigravity, we have seen that the mass of an individual star is not constant outside its radius. Rather, the mass function rises inside the star up to the stellar radius and then it \emph{decays} in the exterior spacetime until it approaches a certain asymptotic value. This then means that the gravitational field of a neutron star in its exterior is not simply given by the typical GR parameterization with a constant mass. A proper analysis of binary pulsar data to place constraints on bigravity would then require that (i) one calculate numerically the dependence of the post-Keplerian parameters on the bigravity coupling constants, and then that (ii) one find the set of bigravity parameters and binary masses that are allowed given the measurements of the post-Keplerian parameters from the timing model.

Measurements of quiescent low mass X-ray binaries can also allow one to infer the radius of neutron stars~\cite{Guillot:2013wu}. The pure hydrogen environment of the neutron star surface producing thermal emission can be modeled which, when combined with an inferred mass from the evolution of the post-Keplerian binary parameters, can be used to infer the radius of the neutron star. Such a measurement can eliminate degeneracies introduced by our ignorance of the equation of state of supranuclear matter in mass-radius relations. In bigravity, the decay of the mass function outside the star will introduce a coupling dependence on the thermal emission models when carrying out the fits to the data.

An extended and precise observation of relativistic binary pulsars may also allow the extraction of the moment of inertia of the stars. This comes about because the timing model is sensitive to the orbital motion of the stars, and the latter depends on the spin angular momentum of the binary components due to spin-orbit precession. Future observations of PSR J0737-3039 may allow for a measurement of the moment of inertia of the primary pulsar to 10\%~\cite{0004-637X-629-2-979}. If such a future measurement is possible, then one could draw an error ellipse in the moment of inertia-mass plane. In GR, this error ellipse will provide a measure of the equation of state of the neutron star. In bigravity, the error ellipse will place a combined constraint on the equation of state and the coupling parameters of the modified theory. 

In addition to the extensions and applications discussed above, another possibility is to consider whether combined gravitational wave and binary pulsar observations can allow for stronger constraints on bigravity. This could be achieved by using the approximately universal relations between the moment of inertia of neutron stars and their Love number found recently in GR~\cite{Yagi:2013bca,Yagi:2013awa,Yagi:2016qmr}. Such tests of bigravity would require the construction of such universal relations in this modified theory, which in turn would require the calculation of neutron stars to second-order in a slow-rotation expansion.

A different avenue to explore is to consider the stability of the stellar solutions found here. This would require the study of perturbations of the numerical solution, which in turn would require the linearization of the modified field equations about these numerical background and an eigenmode analysis for the frequencies of the perturbations. Such an analysis is interesting because of the possibility of super-radiant instabilities in rotating solutions. Indeed, modified theories of gravity with massive fields tend to introduce a potential barrier to perturbations, which then source a super-radiant instability that destroys what at first appears as stationary axisymmetric solutions. If this is the case for massive bigravity for all values of the coupling parameters, then serious doubt would be cast on the theoretical viability of the theory.

\appendix
\section{}
\label{app:Feq}

\begin{align}
\label{eq:tetbasis}
\begin{split}
F_1 &= m^2 \cos^2{\eta} \left[ c_3 U^2 - \left(3 c_3 +1 \right) U\, r + 2 r^2 \left( c_3 + \frac{3}{4} \right) \right], \\
& \mrm{x} \sqrt{\frac{U \left(r-2 m_g \right)}{r \left(U-2 m_f \right)}}, \\
F_2 &= 4 r^2 \pi \rho - \frac{3}{2} m^2 \cos^2{\eta} \bigg[\left(\frac{5}{3} c_3 + 2 \right) r^2 - \left(\frac{8}{3} c_3 + 2 \right) U \, r, \\
& + \left(c_3 + \frac{1}{3}\right) U^2 \bigg], \\
F_3 &= -\frac{1}{2} m^2 \sin^2{\eta} \left[ \left(U-r\right) \left(U - 3 r\right) c_3 + r^2 \right], \\
F_4 &=\frac{m^2 \sin^2{\eta} \left[c_3 U^2 - \left(3 c_3 + 1 \right) U\, r + 2 r^2 \left(c_3 + \frac{3}{4} \right) \right]}{\sqrt{\frac{U \left(r-2 m_g \right)}{r \left(U-2 m_f \right)}}},
\end{split}
\end{align}
The remaining $F_5$, $F_6$, and $F$ are available upon request.

\acknowledgements

N.Y. thanks the hospitality of the Yukawa Institute for Theoretical Physics, where this project was started. We thank Takahiro Tanaka, Paolo Pani, Thomas Sotiriou, and Fawad Hassan, for helpful discussions and comments. We are also grateful for the computational support of the Hyalite High-Performance Computing System, operated and supported by University Information Technology Research Cyberinfrastructure at Montana State University. N.Y. acknowledges support from the NSF CAREER grant PHY-1250636 and NASA grant NNX16AB98G.

\bibliographystyle{apsrev}
\bibliography{RotNSBigravityPaper}
\end{document}